\documentclass[preprint2]{aastex}

\usepackage{epsf}
\usepackage{graphics}

\newcommand{\kms}{km s$^{-1}$}
\newcommand{\cts}{cts s$^{-1}$}
\newcommand{\ax}{$\alpha_{\rm X}$}
\newcommand{\aox}{$\alpha_{\rm ox}$}
\newcommand{\cm}{cm$^{-2}$}
\newcommand{\rb}[1]{\raisebox{1.5ex}[-1.5ex]{#1}}

\newcommand{\pl}{$\pm$}

\shorttitle{RX J2217.9-5941}
\shortauthors{Grupe et al.}

\begin{document}

\def\clipfig#1{\def\lbracket{[}\def\testit{#1}%
    \ifx\testit\lbracket\let\next=\optclipfig\else\let\next=\stdclipfig\fi%
    \next{#1}}
%
\newcommand {\hclipfig} [7] {\clipfig[#7]{#1}{#2}{#3}{#4}{#5}{#6}}
%
\def\usemodepsfig {\global\def\cfmode{x}\typeout{*** set clipfig to PSFIG mode ***}}
\def\usemodeepsf  {\global\def\cfmode{}\typeout{*** set clipfig to EPSF mode ***}}
\def\useunitmm    {\global\def\cfunit{x}\typeout{*** set clipfig to use mm as unit ***}}
\def\useunitcm    {\global\def\cfunit{}\typeout{*** set clipfig to use cm as unit ***}}
\def\clipfigsettings {\ifx\cfmode\empty\def\ccfmode{EPSF }\else\def\ccfmode{PSFIG }\fi%
    \ifx\cfunit\empty\def\ccfunit{cm }\else\def\ccfunit{mm }\fi%
    \typeout{*** current clipfig settings: \ccfmode mode, using \ccfunit as unit ***}}
%
%
%
%
\def\stdclipfig#1#2#3#4#5#6{\ifx\cfmode\empty%
    \let\next=\eclipfig\else\let\next=\pclipfig\fi%
    \next{#1}{#2}{#3}{#4}{#5}{#6}}
\def\optclipfig#1#2]#3#4#5#6#7#8{\ifx\cfmode\empty%
    \let\next=\ehclipfig\else\let\next=\phclipfig\fi%
    \next{#3}{#4}{#5}{#6}{#7}{#8}{#2}}
%
%
%
\newcommand {\pclipfig}[6] {\ifx\cfunit\empty%
        \psfig{figure=#1.ps,width=#2cm,bbllx=#3cm,bblly=#4cm,bburx=#5cm,%
           bbury=#6cm,clip=}\else%
        \psfig{figure=#1.ps,width=#2mm,bbllx=#3mm,bblly=#4mm,bburx=#5mm,%
           bbury=#6mm,clip=}\fi}
\newcommand {\phclipfig}[7] {\ifx\cfunit\empty%
        \hspace{#7cm}\psfig{figure=#1.ps,width=#2cm,bbllx=#3cm,bblly=#4cm,%
           bburx=#5cm,bbury=#6cm,clip=}\else%
        \hspace{#7mm}\psfig{figure=#1.ps,width=#2mm,bbllx=#3mm,bblly=#4mm,%
           bburx=#5mm,bbury=#6mm,clip=}\fi}
%
%
%
\newcommand {\eclipfig}[6]{%
  \ifx\cfunit\empty\epsfxsize=#2cm\else\epsfxsize=#2mm\fi%
  \epsfclipon\epsfverbosetrue%
  \cfcmtopspts{#3}\cfllxi=\cftempi\cfllxf=\cftempf%
  \cfcmtopspts{#4}\cfllyi=\cftempi\cfllyf=\cftempf%
  \cfcmtopspts{#5}\cfurxi=\cftempi\cfurxf=\cftempf%
  \cfcmtopspts{#6}\cfuryi=\cftempi\cfuryf=\cftempf%
  \def\cfstra{\number\cfllxi.\number\cfllxf}%
  \def\cfstrb{\number\cfllyi.\number\cfllyf}%
  \def\cfstrc{\number\cfurxi.\number\cfurxf}%
  \def\cfstrd{\number\cfuryi.\number\cfuryf}%
  \hbox{\epsfbox[{\cfstra} {\cfstrb} {\cfstrc} {\cfstrd}]{#1.ps}}}
\newcommand {\ehclipfig}[7]{%
  \ifx\cfunit\empty\epsfxsize=#2cm\else\epsfxsize=#2mm\fi%
  \epsfclipon\epsfverbosetrue%
  \cfcmtopspts{#3}\cfllxi=\cftempi\cfllxf=\cftempf%
  \cfcmtopspts{#4}\cfllyi=\cftempi\cfllyf=\cftempf%
  \cfcmtopspts{#5}\cfurxi=\cftempi\cfurxf=\cftempf%
  \cfcmtopspts{#6}\cfuryi=\cftempi\cfuryf=\cftempf%
  \def\cfstra{\number\cfllxi.\number\cfllxf}%
  \def\cfstrb{\number\cfllyi.\number\cfllyf}%
  \def\cfstrc{\number\cfurxi.\number\cfurxf}%
  \def\cfstrd{\number\cfuryi.\number\cfuryf}%
  \ifx\cfunit\empty\hspace{#7cm}\else\hspace{#7mm}\fi%
  \hbox{\epsfbox[{\cfstra} {\cfstrb} {\cfstrc} {\cfstrd}]{#1.ps}}%
  \vspace{-1mm}}
%
%
%
\newdimen\cfllxi \newdimen\cfllyi  \newdimen\cfurxi  \newdimen\cfuryi
\newdimen\cfllxf \newdimen\cfllyf  \newdimen\cfurxf  \newdimen\cfuryf
\newdimen\cftemp \newdimen\cftempi \newdimen\cftempf
\newdimen\cfpspoint \cfpspoint=1bp
%
%
%
\newcommand{\cfcmtopspts}[1]{\ifx\cfunit\empty%
  \cftemp=#1cm\else\cftemp=#1mm\fi%
  \multiply\cftemp10\divide\cftemp\cfpspoint%
  \cftempf=\cftemp\divide\cftemp10\cftempi=\cftemp\multiply\cftemp10%
  \advance\cftempf-\cftemp}
%
%
\def\cfmode{}\def\cfunit{}\clipfigsettings
%

\useunitmm

\def \charthoffset {\hspace{0.2cm}} \def \charthsep {\hspace{0.3cm}}
\def \chartvsepcap {\vspace{0.3cm}}
\def \chartvsep {\vspace{0.1cm}}
\newcommand{\putcharta}[1]{\clipfig{#1}{89}{7}{5}{275}{185}}
\newcommand{\putchartb}[1]{\clipfig{#1}{82}{5}{3}{275}{185}}
\newcommand{\chartlineb}[2]{\parbox[t]{18cm}{\noindent\charthoffset\putcharta{#1}\charthsep\putcharta{#2}\chartvsep}}


\title{Chandra Observations of the NLS1 RX J2217.9--5941 \\
}


\author{D. Grupe}
\affil{Astronomy Department, Ohio State University,
    140 W. 18th Ave., Columbus, OH-43210, U.S.A.}

\email{dgrupe@astronomy.ohio-state.edu}

\author{K.M. Leighly}
\affil{Department of Physics \& Astronomy, The University of Oklahoma,
440 W. Brooks St., Norman, OK 73019}

\author{V. Burwitz, P. Predehl}
\affil{MPI f\"ur extraterrestrische Physik, Giessenbachstr. 1, D-85748 Garching,
Germany}

\author{S. Mathur}
\affil{Astronomy Department, Ohio State University,
    140 W. 18th Ave., Columbus, OH-43210, U.S.A.}




\begin{abstract}
We report the results of two Chandra ACIS-S observations from February and
August 2003 of the highly X-ray variable Narrow-Line
Seyfert 1 galaxy RX J2217.9--5941. 
Observations spanning the
   time from the ROSAT All Sky Survey (RASS) through an ASCA observation in
   1998 indicate apparently
   monotonically decreasing flux by a factor of 30.  The
   Chandra observations reveal increased emission over that seen in
   ASCA, supporting a persistent
   variability rather than an X-ray outburst event.
However, the cause of the strong X-ray variability remains unclear.
Our Chandra observations confirm the steep soft X-ray spectrum in the 0.2-2.0
keV band  found during
the ROSAT All-Sky Survey observation (\ax=2.7). The spectral shape of the source
appears to be variable with the spectrum becoming softer when the source becomes
fainter. Best fitting models to the data
include an absorbed broken power law, a blackbody plus power law, 
and a power law with partial
covering absorption. The latter model suggests a variable partial-covering
absorber in the line of sight which can explain in part
the variability seen in RX J2217.9--5941. We suggest that there might be a
population of Narrow Line Seyfert 1 galaxies which are at least at times
highly absorbed. 
\end{abstract}


\section{Introduction}

With the launch of the X-ray satellite ROSAT \citep{tru83} the X-ray energy
range down to 0.1 keV became accessible for the first time. During its half-year
ROSAT All-Sky Survey (RASS, \citet{vog99})
 a large number of sources with steep X-ray
spectra were
detected (\citet{tho98, beu99, schwo00}). About one third to one half
of these sources are AGN. \citet{gru96} and \citet{gru98a, gru04b} 
found that about 50\% of these
bright soft X-ray selected AGN are Narrow-Line Seyfert 1 galaxies (NLS1s). 
These
sources were originally defined by their optical properties (\citet{oster85,
good89}) and they turned out to be the class of AGN with the steepest X-ray
spectra (e.g. 
\citet{stephens89, puch92, bol96, gru96, laor97, bra97, gru98a, lei99b,
vau01, gru04b, wil02}). 
NLS1s are not only AGN with extreme observed properties which are linked
together with the \citet{bor92} Eigenvector 1 relation, they are also 
sources with extreme X-ray variability  (e.g. \citet{bol96, bol97, lei99,
gru04b}.  The most extreme case of X-ray variability found among galactic nuclei
is X-ray transience (e.g. \citet{gezari03, kom02, kom04, 
donley02, vau04, halpern04}) in which a source
appears bright in the X-ray sky only once and becomes very faint by factors 
of up to
several thousand
in later years. 

In the case of X-ray outbursts seen in inactive
galaxies such as RX J1624.9+7554 \citep{gru99a, halpern04}, RX J1242.6--1119 
\citep{kom99b, kom04},
and RX J1420+5334 \citep{grei00} the most plausible explanation is the 
tidal-disruption-of-a-star scenario as suggested by
\citet{frank76} and \citet{rees88}. In the case of the outburst observed in the
Seyfert 2 galaxies IC 3599 \citep{gru95a, bra95} and NGC 5905 \citep{kom99a} the
nuclear activity in these sources suggests that besides the tidal disruption of a
star scenario, it could as well be an instability in the accretion 
disk that caused the
dramatic increase in the X-ray flux. 

The situation is somewhat different in the
NLS1 WPVS 007 \citep{gru95b}. Here
the X-ray flux during the RASS agreed with
that expected from its optical flux suggesting that the high X-ray 
flux seen during the RASS was not caused by an outburst.
Nevertheless, the source almost vanished
from the X-ray sky in later HRI and Chandra observations (\citet{gru01a} and
\citet{vau04}, respectively). A possible explanation of this
behavior is a change in the accretion disk temperature (Grupe et al.
1995b). 

ROSAT and ASCA observations of the NLS1 RX J2217.9--5941 ($\alpha_{2000}$=22h17m56.6s,
$\delta_{2000}=-59^{\circ}41^{'}30.2^{''}$, z=0.160) 
have shown this source is
highly variable in X-rays \citep{gru01a, gru01b}. 
It was rather bright during the
RASS with a mean count rate of
CR=0.83 \cts~in ROSAT's Position Sensitive Proportional Counter (PSPC,
\citet{pfe87}) and showed a decrease in its count rate by a factor of
$\approx$12 between the beginning and the end of its 2 day RASS coverage.
Its X-ray
loudness \aox\footnote{\aox~
is the slope of a hypothetical power-law from 2500 \AA~ to 2 keV;
\aox=0.384$\log~(L_{2500}/L_{2keV})$} of RX J2217.9--5941 during the RASS
agreed well with the value of \citet{yua98} for objects of the 
same luminosity and redshift. After the RASS
the source flux dramatically decreased by factors of about 30 in
later ROSAT High Resolution Imager (HRI) and ASCA observations.
\citet{gru01b} suggested that the X-ray variability 
RX J2217.9--5941 could be explained by a) persistent variability such as
observed in e.g the NLS1 IRAS 13224--3809 \citep{bol97, gallo04c}, 
which is supported
by the \aox~ during the RASS, b) a variable absorber, c) an X-ray outburst such
as seen in IC 3599, or d) that RX J2217.9--5941 is
 an X-ray transient candidate in which a change in the accretion disk
 temperature has shift the X-ray spectrum out of the observed energy window,
 similar to what has been suggested for the NLS1 WPVS 007 \citep{gru95b}.
 In order to test this hypotheses we performed two 5ks ACIS-S
observations with Chandra during guaranteed time observation time granted to 
the Max-Planck-Institute f\"ur extraterrestrische Physik in February and August
2003.

The outline of this paper is as follows: in \S\,\ref{observe} we describe the
Chandra observations and the data reduction, in \S\,\ref{results} we
present the results of the temporal and spectral analysis,
and in \S\,\ref{discuss} we discuss possible origins of the variability
observed in RX J2217.9--5941. 
Throughout the paper spectral indexes are denoted as energy spectral indexes
with
$F_{\nu} \propto \nu^{-\alpha}$. Luminosities are calculated assuming a Hubble
constant of $H_0$ =75 \kms Mpc$^{-1}$ and a deceleration parameter of $q_0$ =
0.0. All errors given in this paper refer to 1$\sigma$ errors.

\section{\label{observe} Observations}

RX J2217.9--5941 was observed by Chandra on 2003-02-23 19:42 - 21:41 (UT) and 
2003-08-16 18:38 - 20:25 (UT) for 4.7 and 5.1 ks with the back-illuminated 
ACIS-S CCD chip S3.  The observations
were performed in the Faint Mode with the standard frame time of 3.2s. 
Source photons were collected 
in a circular region with a radius r=3$^{''}$
and background photons in an annulus with an inner radius r=4.3$^{''}$ and an
outer radius r=12$^{''}$.
Spectra were extracted from the primary event files with CIAO version 3.0.2 
and analyzed using XSPEC 11.2. The calibration database used was CALDB version
2.25. The response matrices and auxiliary response file for the effective area
were created with the CIAO tasks {\it mkrmf} and {\it mkarf}.
The effective areas were corrected by the task {\it apply\_acis\_corr} in
order to correct for contamination of a lubricant on the ACIS entrance window.
The spectroscopic data were rebinned by {\it GRPPHA} to have least 20 photons per
bin.
For count rate conversion between the different X-ray missions, PIMMS 3.2
was used.

\section{\label{results} Results}

\subsection{Source position}
Chandra's superior pointing accuracy and its spatial resolution enables us to
measure the most accurate source position compared with other X-ray missions. From
the February 2003 observation of RX J2217.9--5941 we measured the 
source positions to be
 $\alpha_{2000}=22^{\rm h}17^{\rm m}56.63^{\rm s}$ and
$\delta_{2000}=-59^{\circ}41^{'}31.2^{''}$ consistent within 1$^{''}$ of the
results of our HRI observation  \citep{gru01b}.

\subsection{\label{variability} X-ray variability}

\begin{figure}[h]
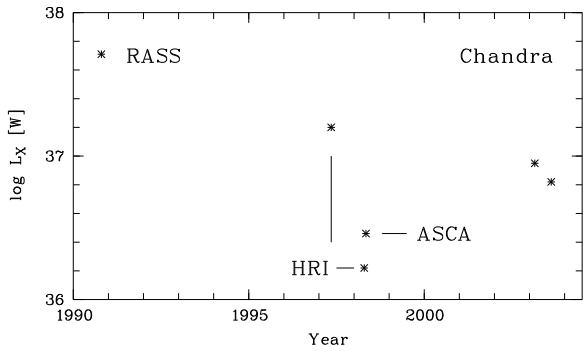

\clipfig{DGrupe.fig1}{80}{20}{45}{280}{200}
\caption[ ]{\label{rxj2217_loglx_light} Long-term light-curve of the 0.2-2.0
rest-frame X-ray luminosity of RX J2217.9--5941
}
\end{figure}

The mean ACIS-S count rate in the 0.5-7.2 keV band during the February 2003
observation was CR=0.1258\pl0.0052 \cts, and 0.0500\pl0.0031 \cts~in the 0.5-3.6
keV range during the August 2003 observation.
Figure\,\ref{rxj2217_loglx_light} displays the long-term lightcurve of RX
J2217.9--5941 including all X-ray observations performed on this source. 
Because the HRI observations do not provide spectral information and the ASCA
spectral data have large uncertainties \citep{gru01b},
the 0.2-2.0 rest frame luminosities of the HRI and ASCA observations
were derived by using PIMMS assuming the X-ray spectral index of \ax=2.7 
and no significant spectral changes. 
The unabsorbed rest-frame 0.2-2.0
X-ray luminosities of the Chandra ACIS-S observations of log $L_{\rm
0.2-2.0~keV}$ = 36.95 [W] and 36.82 [W] (43.95 and 43.82 [ergs s$^{-1}$])
of the February and August 2003
observations, respectively.

We used the hardness ratio to look for spectral variability between the
observations.  For our purpose we
define the hardness ratio as HR=(H-S)/H+S) with the counts S in the 
soft band in the 0.5-1.0 keV
and hard counts H 
in the 1.0-2.0 keV energy ranges. We used  this particular
definition to avoid any contribution in the February 2003 spectrum
from a possible mis-calibration around 2.3 keV (see \S3.3). Using this
definition of the hardness ratio we found HR=--0.205\pl0.081 
and HR=--0.492\pl0.092 during the February and August 2003 observation,
respectively,
suggesting a change in the X-ray spectrum in about half a year. The spectrum
became softer with decreasing count rate.  Note that also pileup
causes the raw spectrum to become flatter. Because the February 2003 data are
affected by pileup and the August data are not this could cause in part the
variability we observe. 

\subsection{\label{xspectra} Spectral analysis}

At first we fitted an
 XSPEC power law model only to the data in the 0.5-2.0 observed energy
range to be consistent with the ROSAT PSPC range. The results of these power law
fits are listed in Table\,\ref{rxj2217_spec_tab}. In the February 
2003 observation, the XSPEC {\it PILEUP} model of \citet{davis01} is
needed to model the spectrum correctly. 
This model takes into account the fact that the count
   rate of this object is sufficiently high that more than one photon
   is obtained in a single pixel during the nominal frame time of 3.2
   seconds.
 This causes a change in the grades  by
a grade morphing parameter $\alpha$ which determines the number of
'good grades' (see \citet{davis01} for details) by {\it number of good grades} =
$\alpha^{\rm (P-1)}$ where P is the number of piled up photons. 
For the August 2003 observation the count rate was low enough not to be
significantly affected by pileup.
The February and August 2003 data spectra are well-fitted by a simple
power law with Galactic absorption (2.58$\times 10^{20}$\cm; \citet{dic90}). 
The results of the spectral fits are listed in 
 Table\,\ref{rxj2217_spec_tab}. Both
observations confirm the steep X-ray spectrum of \ax=2.7 that was found during
the RASS observation \citep{gru01b}.

\begin{figure*}[h]
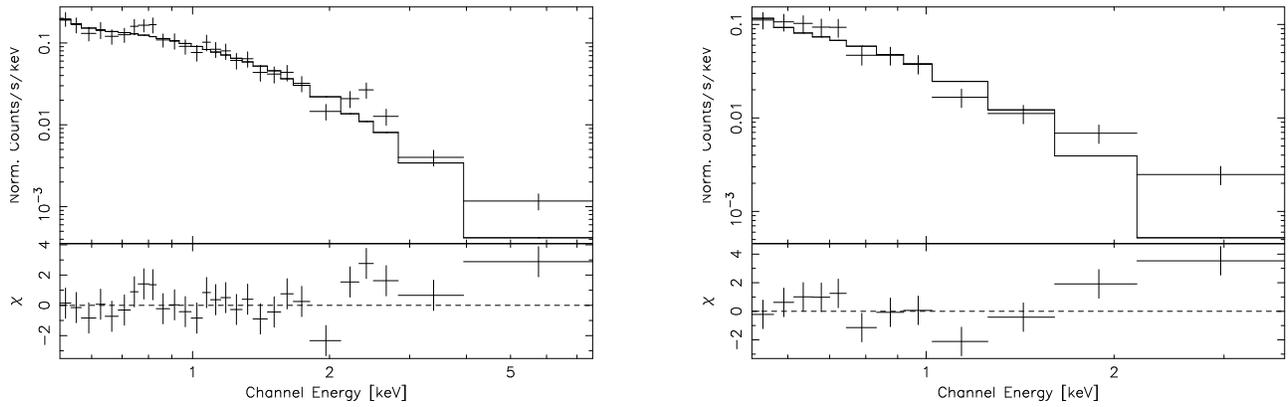

\chartlineb{DGrupe.fig2a}{DGrupe.fig2b}
\caption[ ]{\label{rxj2217_plot_05_72} power law fit with Galactic absorption to
the February 2003 (left) and August 2003 (right) spectra of RX J2217.9-5941. The
spectral slope was fixed to the values determined in the 0.5-2.0 keV range. 
(Table\,\ref{rxj2217_spec_tab}).
Notice the flattening of spectra towards harder X-ray energies and 
the strong residuals around 2.3 keV in the February 2003 spectrum.
}
\end{figure*}

The spectra seem to become more complicated towards higher energies
(Figure\,\ref{rxj2217_plot_05_72}). However, the February 2003 spectrum is, even
though it is not a good fit, still consistent with a simple power law model. On
the other hand, the August 2003 spectrum is not
(Table\,\ref{rxj2217_spec_tab}). The spectra were fitted with power law models
with the X-ray spectral index left free and fixed to the value determined in the
0.5-2.0 keV range. In all cases the spectra show a flattening towards higher
X-ray energies. In a first approach to fit these spectra we used a broken
power law with the absorption parameter fixed to the Galactic value.
This model results in acceptable fits
with a soft X-ray spectral index of \ax=2.58\pl0.61 and 3.55\pl0.30 for the
February and August 2003 data, respectively (Table\,\ref{rxj2217_spec_tab}).
 This model
suggests a soft X-ray excess above a hard X-ray spectrum with a flatter spectral
index of \ax$\approx$1.0. 

A partial-covering absorber can also potentially
    explain a flattening of the spectrum toward higher energies.
 This model has been successfully applied to the XMM
data of several NLS1s (e.g. \citet{bol02, bol03} and \citet{gru04c}). Applying
a partial-covering absorber to the spectra of RX J2217.9--5941 also gives
acceptable fits. A partial-covering absorber is also able to explain the
variability observed in RX J2217.9--5941.
However, one has to keep in mind that the quality especially 
of the August 2003 data do not constrain of the partial-covering
absorber parameters, particularly when the X-ray slope is free to vary.
The X-ray spectra from NLS1s
are also frequently parameterized using a blackbody plus
power law spectrum (e.g. \citet{lei99}). However, this model does not improve
the fits to the data (Table\,\ref{rxj2217_spec_tab}).
The February 2003 data (Figure\,\ref{rxj2217_plot_05_72}) also
show strong residuals
around 2.3 keV. This feature can be modeled by a Gaussian line which
significantly improves the fit.  However, the
existence of a real emission line at this energy is rather uncertain.
\citet{aldcroft03} reported a similar feature in their Chandra ACIS-S
spectrum of 3C212 and concluded that the 'line' is most likely instrumental
nature. 

As mentioned in \S\,\ref{variability} there seems to be a change in the hardness
ratios between the February and August data. In order to check whether this is
real or just the effects of pileup we used the simultaneous fit to both data
sets in XSPEC by applying a blackbody plus power law fit to the data as given in
Table\,\ref{rxj2217_spec_tab}. The blackbody temperature and the power law index
were fixed and the normalizations
were left free. The changes in the normalization agree with results from the
changes in the hardness ratios: While the normalization of the blackbody of the
soft X-ray part decreases by a factor of about 2 between February and August
2003, the normalization of the power law decreased by a factor of about 5, 
suggesting also that the
spectrum became softer when the source it became fainter. 

\section{\label{discuss} Discussion}
The main aspect of monitoring the NLS1 RX J2217.9--5941 with Chandra was to
investigate whether the source is an X-ray transient AGN, 
such as WPVS 007 \citep{gru95b},
or if it is simply a highly variable source that was observed in a bright state
during the RASS and in fainter states in later ROSAT HRI and ASCA observations.
Our new Chandra ACIS-S observations favor the latter scenario and support 
one of the
speculations of \citet{gru01b} that the variability found in RX J2217.9--5941 is
most likely persistent.
There is no
indication from the Chandra observations that RX J2217.9--5941 is an X-ray
transient AGN.  The X-ray
loudness  of RX J2217.9--5941 during the RASS was
\aox=1.52 \citep{gru01b} and agrees perfectly with the \aox~given by
\citet{yua98} for radio-quiet AGN with similar redshifts and luminosities as RX
J2217.9--5941 which also
supports the persistent variability nature of the source. This \aox~suggests also
that the X-ray flux measured during the RASS is the 'normal', expected flux.
Compared with the RASS
observation, during the HRI and ASCA observations the optical to X-ray slope
dropped to \aox=2.0. The optical spectra of RX J2217.9--5941 taken between 1992
and 1998 \citep{gru01b} do not suggest any optical variability.
The remaining question about RX J2217.9--5941 is what causes the X-ray
variability. There are two possible explanations, 1) intrinsic variability or 2)
a variable cold absorber in the line of sight.

We also found in
\S\,\ref{variability} that RX J2217.9--5941 is
not only variable in flux, but also in its spectral shape. While most AGN
become harder with decreasing flux (e.g. \citet{gallo04a, dewangan02, lee00,
chiang00}) in RX J2217.9--5941 the source becomes softer when the flux
decreases.
Even though this is unusual, it is not unobserved in NLS1s: e.g. PKS 0558--504
\citep{mario01}, 
RX J0134.2--4258 \citep{gru00} or 1H 0707--495 \citep{gallo04b, fabian04}. 
In the case of the radio-loud object
PKS 0558--504 \citet{mario01} discussed the observed spectral variability in the
contest of a contribution of a jet. This model can not be applied for RX
J2217.9--5941, which is a radio-quiet source \citep{gru01b}. 
In the case of RX
J0134.2--4258 \citet{gru00} 
excluded a warm absorber as the explanation for these 
findings\footnote{Note that \citet{kom00} discussed the variability in RX J0134.2--4258
in the context of the presence of a warm absorber, however we did not find any
evidence in the ASCA spectrum of RX J0134.2--4258 for such an absorber.},
but discussed  a change in the accretion disk corona as a  possible explanation.
Recently \citet{wang04} found that the corona becomes weak when the Eddington
ratio increases. As a result the hard X-ray flux decreases with regards to the
bolometric luminosity. In the case of 1H 0707--495, \citet{fabian04} explained
the variability by ionized disk reflection.
Both models may also explain the spectral behavior of RX J2217.9--5941, but for
the current data it is not possible to draw  final conclusions.

As an alternative 
the source can be intrinsically  absorbed  
and this absorber is variable. It has been suggested by e.g. \citet{abra00} 
that clouds of neutral matter in
the line of sight may cause the variability observed in X-ray in AGN.
Such cold absorber has been suggested by e.g. \citet{mckernan98} and
\citet{lamer03} to explain the X-ray variability of MCG--6--30--15 and NGC 3227.
A simple cold absorber of neutral gas in RX J2217.9--5941
can cause the count rate to decrease as
observed between the RASS and the pointed HRI observation in 1997 and 1998.
Assuming the intrinsic spectrum did not change, it requires an absorption
column of neutral gas of 8$\times~10^{20}$\cm~ to account for the HRI count
rate observed in 1997 and a column density of about 9$\times~10^{21}$\cm~
to explain the second HRI observation assuming the intrinsic luminosity has not
changed.

Most likely the absorber in RX J2217.9--5941 is a partial-covering
absorber that blocks most of the X-rays but has 'leaks' either through
parts of much
lower column density in the absorber or by scattering (e.g. \citet{gru04c}). In
the case of 1H 0707--495 \citet{bol02} and \citet{tanaka04} explained the X-ray
spectrum by a partial-covering absorber which can explain the spectral
variability in this source \citep{gallo04b} which is similar to what we found in
RX J2217.9--5941. 
Even though we can not put constraints on the absorber
parameters the 0.5-7.2 February 2003 and 0.5-3.5 keV August 2003
data can be fitted by a power law model with
partial-covering absorber,
suggesting that the variability at least in part maybe due to a variable partial
covering absorber in the line of sight. 
 When using  partial-covering models as listed in
Table\,\ref{rxj2217_spec_tab} the unabsorbed fluxes in both observations are
practically the same with rest frame fluxes of log $F_{\rm 0.5-3.5 keV}$=--14.2
[W m$^{-2}$] (--11.2 [ergs s$^{-1}$ cm$^{-2}$]).
A partial-covering absorber may also be responsible for the X-ray
transience found in the NLS1 WPVS 007 and might explain why no hard X-ray
photons have been observed in this peculiar source \citep{gru95b}. This raises
the question if there is a population of NLS1s which is at least at times
highly absorbed in X-rays.
This hypothesis is supported by the findings of \citet{wil02} who
studied
150 NLS1s selected from the
Sloan Digital Sky Survey (SDSS, \citet{york00}) and found that 10-20
sources were not detected by ROSAT. Their optical spectra did not show any
differences to those of X-ray selected NLS1s.
 From Chandra
follow-up observations of 17 of these sources, \citet{wil04} suggest that there
is a population of NLS1s which are either 
intrinsically X-ray weak or are absorbed. 
The composite
spectrum of \citet{wil04} does not show signs of intrinsic absorption, but
the quality of these snap-shot
Chandra observations did not allow any detailed spectral analysis in particular
to check for the presence of a partial-covering absorber.

\acknowledgments

We would like to thank Himel Gosh, Stefanie Komossa, and Luigi Gallo
for comments and suggestions on the
manuscript. We would also like to thank the anonymous referee for a fast and
constructive referee's report that significantly improved the paper.
This research 
has made use of the NASA/IPAC Extra-galactic
Database (NED) which is operated by the Jet Propulsion Laboratory, Caltech,
under contract with the National Aeronautics and Space Administration. 
The ROSAT project was supported by the Bundesministerium f\"ur Bildung
und  Forschung (BMBF/DLR) and the Max-Planck-Society. This work was supported in
part by NASA grant NAG5-9937.


\begin{deluxetable}{llccccccccr}
\tabletypesize{\scriptsize}
\tablecaption{Spectral Fit parameters to the Chandra ACIS-S 
data of RX J2217.9--5941. The February 2003 data were corrected for pile up. 
 \label{rxj2217_spec_tab}}
\tablewidth{0pt}
\tablehead{
& & \colhead{E-range} &  & \colhead{$\rm kT_{bb}$} 
&  \colhead{Pileup} & \colhead{$E_{\rm break}$} & & 
\colhead{$N_{\rm H,pc}$} \\
\colhead{\rb{Obs.}} &
\colhead{\rb{Model}} & \colhead{keV}    &
\colhead{\rb{$\alpha_{\rm X}$}} & \colhead{keV} &
\colhead{\rb{$\alpha$}} &
\colhead{keV} & \colhead{\rb{$\alpha_{\rm X,h}$}} & 
\colhead{$10^{22}$\cm} & \colhead{\rb{f$_{\rm pc}$}} & 
\colhead{\rb{$\chi^2$/DOF}} 
} 
\startdata
02/03 & po\tablenotemark{1}  
& 0.5-2.0 & 2.50\pl0.65 & --- & 0.66\pl0.19 & --- & --- & --- & --- & 9.9/19 \\
& & 0.5-7.2 & 2.24\pl0.34 & --- & 0.63\pl0.14 & --- & --- & --- & --- & 33.3/25 \\
& zpcf po\tablenotemark{2} & 
0.5-7.2 & 2.50 (fix) & --- & 0.63 (fix) & --- & --- & 
15.2\pl7.5 & 0.80\pl0.03 & 27.7/25 \\ 
& bknpo\tablenotemark{3} & 0.5-7.2 & 2.56\pl0.61 & --- & 0.66\pl0.14 &
1.97\pl0.88 & 0.94\pl1.02 & --- & --- & 26.9/23 \\ 
& bb po\tablenotemark{4} & 0.5-7.2 & --- & 0.101\pl0.016 & 0.68\pl0.26 & --- & 1.50\pl0.27
& --- & --- & 31.3/23 
\\ \\
08/03 & po\tablenotemark{1} &
0.5-2.0 & 3.38\pl0.26 & --- & --- & --- & --- & --- & --- & 6.9/8 \\
&  &  0.5-3.6 & 2.95\pl0.29 & --- & --- & --- & --- & --- & --- & 28.6/10 \\
& zpcf po\tablenotemark{2} & 0.5-3.6 & 3.00 (fix) & --- & --- & --- & --- & 
6.98\pl2.20 & 0.94\pl0.03 & 9.3/9 \\
& bknpo\tablenotemark{3} & 0.5-3.6 & 3.55\pl0.30 & --- & --- & 1.25\pl0.12 &
0.32\pl0.44 & --- & --- & 5.5/8 \\
& bb po\tablenotemark{4} & 0.5-3.6 & --- & 0.098\pl0.008 & --- & --- &
0.23\pl0.42 & --- & --- & 4.8/8 
\enddata

\tablenotetext{1}{power law model with Galactic absorption fixed to 2.58
10$^{20}$\cm~\citet{dic90} }
\tablenotetext{2}{power law model with Galactic absorption fixed to 2.58
10$^{20}$\cm~\citet{dic90} and partial-covering absorption}
\tablenotetext{3}{broken power law model with Galactic absorption fixed to 2.58
10$^{20}$\cm~\citet{dic90} }
\tablenotetext{4}{blackbody plus power law model with galactic absorption 
fixed to 2.58 10$^{20}$\cm~\citet{dic90} }
\end{deluxetable}

\end{document}